# Efficient Absorption of Terahertz Radiation in Graphene Polymer Composites

Zahra Barani[1,2,*] , Kamil Stelmaszczyk[3,*], Fariborz Kargar[1,2], Yevhen Yashchyshyn[3,4], Grzegorz Cywiński[3], Sergey Rumyantsev[3] and Alexander A. Balandin[1,2,†]

[1]Nano-Device Laboratory, Department of Electrical and Computer Engineering, University of California, Riverside, California 92521 USA

[2]Phonon Optimized Engineered Materials Center, University of California, Riverside, California 92521 USA

[3]CENTERA Laboratories, Institute of High-Pressure Physics, Polish Academy of Sciences, Warsaw 01-142 Poland

[4]Institute of Radioelectronics and Multimedia Technology, Warsaw University of Technology, Warsaw 00-665 Poland

[*] Contributed equally to the work.
[†] Corresponding author (A.A.B.): balandin@ece.ucr.edu ; web-site:  http://balandingroup.ucr.edu/

**ABSTRACT:** We demonstrate that polymer composites with a low loading of graphene, below 1.2 wt. %, are efficient as electromagnetic absorbers in the THz frequency range. The epoxy-based graphene composites were tested at frequencies from 0.25 THz to 4 THz, revealing total shielding effectiveness of 85 dB (1 mm thickness) with graphene loading of 1.2 wt% at the frequency $f$=1.6 THz. The THz radiation is mostly blocked by absorption rather than reflection. The efficiency of the THz radiation shielding by the lightweight, electrically insulating composites, increases with increasing frequency. Our results suggest that even the thin-film or spray coatings of graphene composites with thickness in the few-hundred-micrometer range can be sufficient for blocking THz radiation in many practical applications.

# ■ INTRODUCTION

The shielding of the electromagnetic (EM) energy in the Terahertz (THz) frequency range is important for both reducing EM interference among various devices and protecting humans.[1–3] At the same time, absorbing EM radiation in the THz band can be particularly challenging. There are numerous requirements imposed on the materials used for EM interference (EMI) shielding, such as thickness of the coating layer, weight limits, mechanical and thermal properties, electrical isolation or conductance. Absorbing EM energy rather than reflecting it back to the environment is often required for commercial, *e.g.,* 6G technology, and defense applications.[4] Many existing EMI shielding materials for the high frequency bands, including metallic coatings, mostly redirect the EM energy by reflection owing to their high electrical conductivity. The reflection protects electronic components but, at the same time, may negatively affect human health.[2,3]

A new promising approach for EMI shielding is the use of polymer-based materials with electrically conductive fillers.[5–7] Recent studies reported the use of carbon fibers,[8,9] carbon black,[9,10] bulk graphite,[11,12] carbon nanotubes,[13–16] reduced graphene oxide,[17–26] graphene[27–30] and, combination of carbon allotropes with other particles.[24,26,27,29,31–36] Graphene was used successfully as the filler in composites, which were tested in the MHz and GHz frequency ranges.[23,27–30,37–39] There were two reports on the use of graphene composites in the sub-THz range.[40,41] Experimental and theoretical studies with individual graphene layers and graphene meta-surfaces suggest that graphene interacts efficiently with EM radiation in the THz range.[42–46] Available data suggest that graphene particularly well absorbs the radiation rather than reflects it in the high GHz frequency range.[41] However, up to date we are not aware of any report of testing composites with graphene in the THz range.

Here, we report the results of testing of the epoxy composites with the low loading of graphene, below 1.2 wt. %. The term graphene, in the context of composite studies, is used to identify a mixture of single layer graphene (SLG) and few layer graphene (FLG), with the thickness approximately below 20 atomic planes, and the micrometer-scale lateral dimensions. It is found that such graphene composites are efficient as electromagnetic absorbers in the THz frequency range. The THz radiation is mostly blocked by absorption rather than reflection. The efficiency of the THz radiation shielding by the lightweight, electrically insulating composites, increases with increasing frequency. Our results suggest that even the thin-film or spray coatings of graphene composites with thickness in the few-hundred-micrometer range can be sufficient for blocking THz radiation in many practical applications.

■ **EXPERIMENTAL SECTION**

**Materials:** For this work, we utilized commercially available graphene (xGnP®H-25, XG-Sciences, U.S.A.) to prepare the composites. The samples were prepared in the form of disks with a diameter of 25.6 mm and thicknesses from 0.9 mm to 1.0 mm. The sample thickness affects the total absorption and the total shielding efficiency of the composites. A scanning electron microscopy (SEM) image of graphene fillers and an optical photo of pristine epoxy and a representative graphene composite are shown in Figure 1.

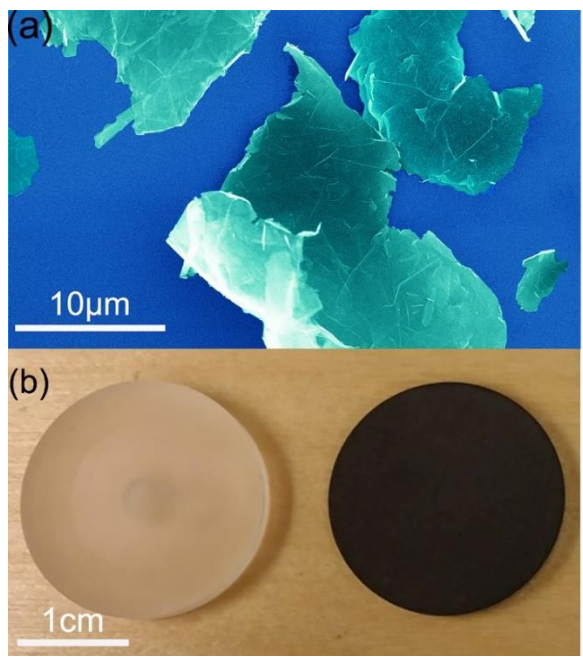

**Figure 1:** (a) Scanning electron microscopy image of the graphene fillers with the average lateral dimension of ~25 μm used in the composite preparation. (b) Optical image of the pristine epoxy (left) and epoxy with 1 wt. % of graphene (right).

**THz Measurement Setup:** The EMI shielding efficiency of the epoxy-based graphene composites were investigated using the Terahertz-Time Domain Spectroscopy (THz-TDS) system (Toptica Photonics AG).[47–49] A train of short THz pulses of ~1 ps duration and the repetition rate of 100 MHz was sent through the sample pellets made of graphene-based composites or was reflected from their surfaces depending on the experimental geometry. The high-mobility InGaAs photoconductive antennas were used as both transmitter (Tx) and receiver (Rx). The antennas were fed with an all-fiber pulsed laser centered at 1.56 μm wavelength. As opposed to 1 ps duration of the THz pulses, the duration of the laser pulses was much shorter, $\tau \sim 80\ fs$. Owing to this property a time narrow fraction of the THz pulse could be sampled at one given delay of the laser pulse. To retrieve temporal profiles of the pulses from the THz pulse train a variable delay stage scanned the THz wave packet with the laser probe pulse. The high repetition rate of the pulse train allowed to apply averaging over $1000 - 2000$ pulses during a single minute duration of an individual measurement. A Fast Fourier Transform (FFT) algorithm was used to calculate the amplitudes and phases of the transmitted and reflected fields as a function of frequency.

A schematic of the THz-TDS experimental setup is shown in the Figure 2. In addition to the laser part, it includes the THz beam delivery part based on the so-called $4f$ optical arrangement,[50] consisting of four 90° off-axis parabolic mirrors. The two of these mirrors, applied to compensate for the divergence of the beam, were located, after the emitter and, after the sample. The other two provided focusing before the pellet and detector. The focal lengths of the mirrors were chosen to ensure symmetrical propagation on the distance from the radiation source to the sample and from the sample to the detector. The diameter of the mirrors was 25.4 mm, and the shorter and longer focal lengths were 50.8 mm and 101.6 mm, respectively. Using these values one can estimate that the Rayleigh lengths $z_R = \pi \omega_0^2 / \lambda$ ($\omega_0-$ beam waist in focus, $\lambda = c/f$, where $c-$ speed of light and $f-$ radiation frequency) of the pulse frequency components ($0.25 - 4$ THz) were not shorter than ~1.5 mm, *i.e.,* substantially longer than the thickness of the composite pellet. Such situation typically indicates a good collimation of the beam near the focal spot, providing equal absorption lengths for all frequency components. Note that the setup shown in Figure 2 was used only for the measurements of the transmission coefficients, $T$. To measure the reflection coefficients, $R$, the receiver was moved to be on the same side of the sample as the emitter, and the amplitude of the EM radiation reflected from the surface of the pellet was measured. A special care was taken to deliver the beam at nearly normal incidence with respect to the sample, resulting in a small deviation of only 4° from the vertical.

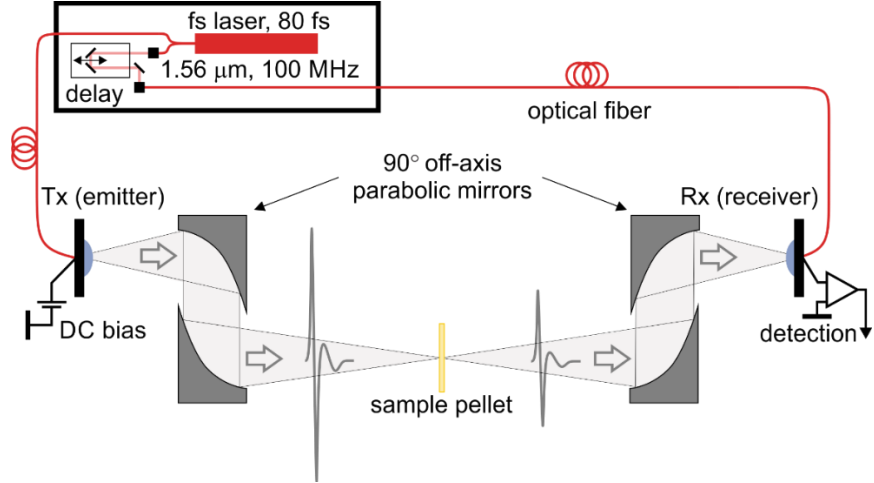

**Figure 2:** Schematic of the experimental setup used for the measurements in the high-frequency range from 250 GHz to 4 THz.

## ▪ RESULTS AND DISCUSSION

The standard procedure involving measurements of the transmission, $T$, and reflection, $R$, coefficients commonly used in the $S$-matrix method[14] was applied to determine the absorption properties of composites. The transmission coefficient of composites was calculated via measuring the amplitude of the electric field transmitted through the sample, $\hat{E}_{21s}$, and through the reference field, $\hat{E}_{21e}$. The latter is measured in an empty space after the removal of the composite. On the basis of these two measurements $T$ was calculated as:

$$T = |S_{21}|^2 = \left|\hat{E}_{21s}/\hat{E}_{21e}\right|^2. \tag{1}$$

Similarly, the reflection coefficient was determined by measuring the electric field, $\hat{E}_{11s}$, reflected from the surface of the epoxy composite. The reference field, $\hat{E}_{11s}$, was measured with the THz pulse reflected from the polished metallic plate, which is a good approximate of an ideal reflector in the THz band. An equivalent of Eq. (1) was used to calculate $R$ as follows:

$$R = |S_{11}|^2 = \left|\hat{E}_{11s}/\hat{E}_{11m}\right|^2. \tag{2}$$

The electric fields from Eqs. (1) and (2) and the phase relations required to calculate the amplitude ratios were provided by the FFT. Finally, to determine the absorption characteristics of EMI shielding material, the measured values of $T$ and $R$ were used to calculate the effective absorption coefficient, $A_{eff}$, defined as:

$$A_{eff} = (1 - R - T)/(1 - R). \tag{3}$$

Compared with the standard definition of the absorption coefficient $A = 1 - R - T$, the modified definition describes the actual absorption of the composite material accounting for the fact that some part of the incident EM energy is always reflected from the surface of the sample.

The total shielding effectiveness, $SE_T$, and shielding *via* different mechanisms of reflection, $SE_R$, and absorption, $SE_A$, can be calculated knowing $R$ and $A_{eff}$ as follows:[41]

$$SE_R = -10\log(1 - R), \tag{4}$$

$$SE_A = -10\log(1 - A_{eff}), \tag{5}$$

$$SE_T = SE_R + SE_A. \tag{6}$$

Figure 3 (a) shows the reflection coefficient, $R$, for the pristine epoxy and three epoxy-based composites with the graphene loading ranging from 0.8 wt % to 1.2 wt%. The data are presented in the frequency range from 0.25 THz to 4 THz. The oscillations of the reflectivity at low frequencies, are due to the multiple reflections from the sample's front and back surfaces, causing Fabry–Pérot interference features. The sharp spikes at $f > 1.5$ THz are due to the absorption by the water molecules in the atmosphere. Within the whole frequency range, the reflection from the epoxy-based composites is small. Figure 3 (b) presents the reflection shielding effectiveness of samples as a function of frequency. One can see that the shielding by reflection in the whole frequency range is below 1 dB.

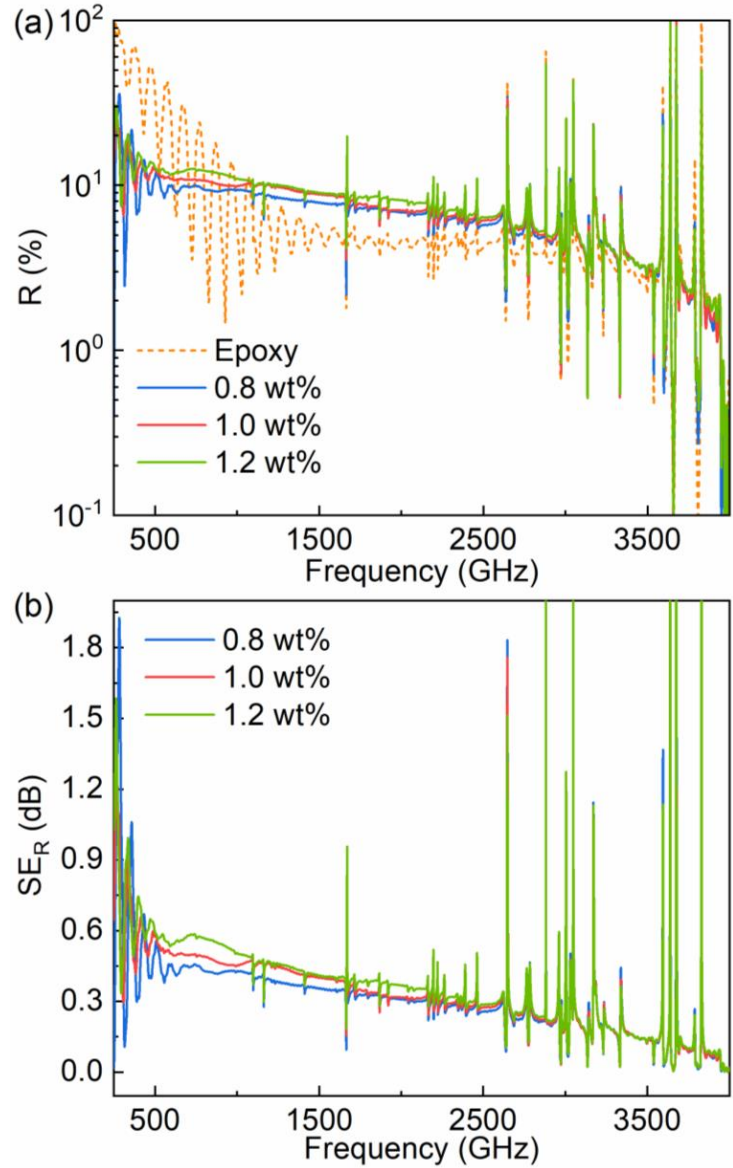

**Figure 3:** (a) Reflection coefficient, $R$, and (b) reflection shielding effectiveness, $SE_R$, of the pristine epoxy and the composites with a low loading of graphene as a function of frequency. Note that the shielding by reflection is negligible in all composites. The experiments have been conducted using an aluminum plate as a reference.

The transmission coefficient, $T$, for the same set of samples is shown in Figure 4 (a). It is interesting to note that the transmission decreases with the increasing frequency. For composites with graphene loading of 1.2 wt. %, and at $f = 1.6$ THz, $T \sim 10^{-7}$%, confirming that almost all the incident EM wave is blocked by the composites. This level of the measured signal is already close to the sensitivity limit of the measurements. As expected, the higher the graphene loading, the smaller is the transmission. The exception is only observed at the frequencies below $f = 0.4$ THz, where the transmission spectra for the composites with 1.0 wt% and 1.2 wt% loading fractions of graphene coincide. The low reflection and low transmission of the epoxy-based composites in the THz frequency band show that this material effectively shields EM radiation mainly due to the absorption. It is important to note that composites with these low graphene loadings are electrically insulating. Therefore, they can be used as adhesives and environment protective layers for the circuits containing multiple electronic components without shortening them. The dashed lines in Figure 4 (a) show the result of recalculation of the experimental data for 1-mm thick sample to the thickness of 200 μm. It is seen that even these thin layers can protect effectively at the THz frequency band. The total shielding effectiveness of the composites, $SE_T$, is presented in Figure 4 (b). One can see that 1.2 wt.% graphene composites provide total shielding of ~80 dB at $f$~1.6 THz, which is more than sufficient for many industrial applications. This result suggests that graphene composite absorbers with small thicknesses can be deposited as a protective paint, *e.g.* spray coated.

It is interesting to note that the slopes and the values of the transmission coefficient, $T$, and the shielding effectiveness, $SE_T$, are nearly the same for the composites containing 1.0 wt.% and 1.2 wt. % graphene at the frequencies below $f = 0.4$ THz. On the other hand, the slopes of $SE_T$ curves corresponding to the samples containing 1.0 wt. % and 0.8 wt. % of graphene are the same for the frequencies $f > 0.4$ THz. Similar trends can be observed for the curves representing the shielding effectiveness related to the absorption, $SE_A$ (see Figure 5 (b)). Since the slope of the $SE_A$ curve is typically a good estimate of the high-frequency resistivity component of a composite material, $\rho$,[41,51] we speculate that the observed change in the $SE_A$ curve slope with the frequency indicates the frequency-dependent percolation effect. To the best of our knowledge, this is the first report of the percolation effect in composites in the TH frequency range.

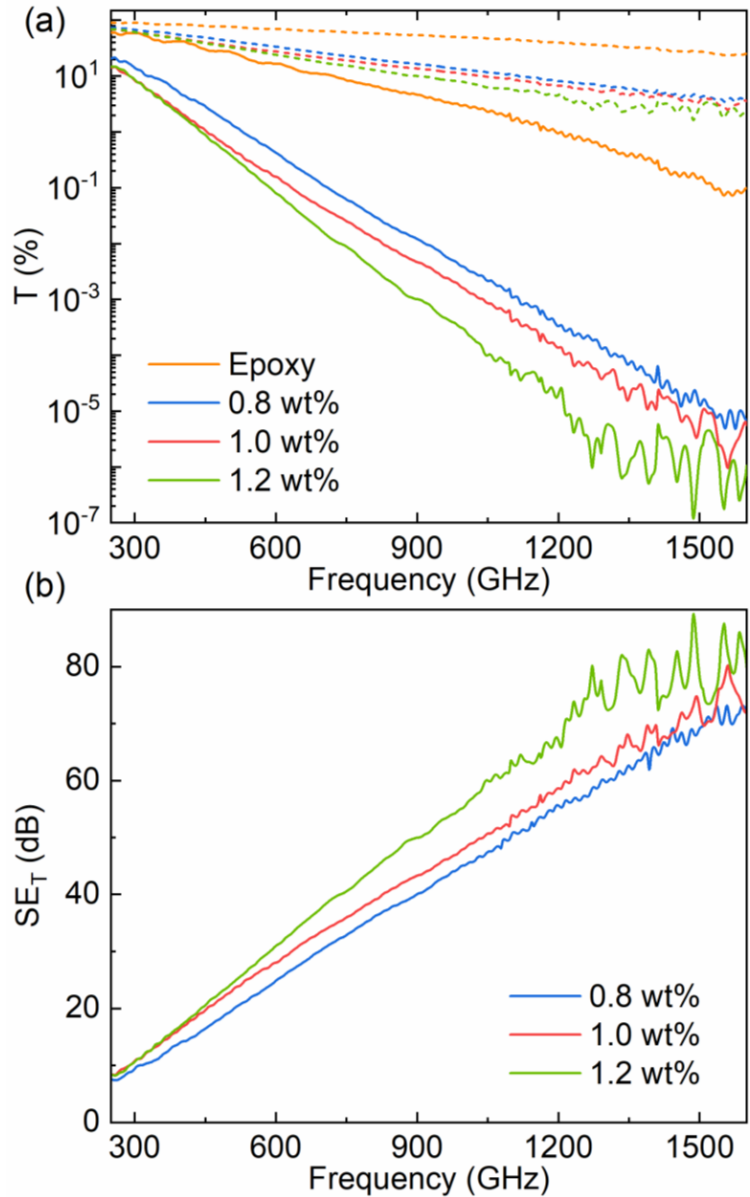

**Figure 4:** (a) Transmission coefficient, $T$, and (b) total shielding effectiveness, $SE_T$ of the composites with a low loading of graphene as a function of frequency. Note that $SE_T$ increases linearly with the frequency. For $f \geq 1600$ GHz, the shielding effectiveness of the samples increases beyond the measureable upper limit of the instrument. The solid lines are the experimental results for samples with 1 mm thickness. The dashed lines are the calculated $T$ and $SE_T$ for 200 μm thick samples, provided for comparison. The change in the slopes of the curves can be associated with the high-frequency percolation effect in the composites.

The frequency dependence of $A_{eff}$ and $SE_A$ are shown in Figure 5 (a-b). The data demonstrate that absorption is the dominant mechanism for blocking the EM waves in the graphene-based polymer composites. One can see that especially at frequencies $f > 0.6$ THz, these composites absorb EM radiation almost completely. The composites with $1.2$ wt. % of graphene provide $\sim 45$ dB and $\sim 80$ dB shielding effectiveness by absorption at $f \sim 0.8$ THz and $f = 1.6$ THz, respectively (Figure 5 (b)). These samples reveal substantially enhanced absorption shielding effectiveness comparing to PDMS composites with $3$ wt. % graphene content in the near-THz frequency range .[40] The extraordinary absorptive properties of graphene composites in the near-THz frequency range was mostly attributed to the activated interaction of atomic vibrations of polymer matrix with the $\pi$-band to polaron band transitions in graphene in the THz frequency range.[52] This interaction, which is in resonance with THz radiation causes strong absorptive properties, yielding a significant THz-range EM shielding behavior.[52] Shielding by absorption is crucial in applications where the EM reflection is undesired. Absorption shielding characteristics of graphene-based composites surpass the most conventional polymeric composites with high loading of ceramic fillers, such as $BaTiO_3$. Ceramic fillers are usually used in composites to enhance the absorption component of total shielding.[53]

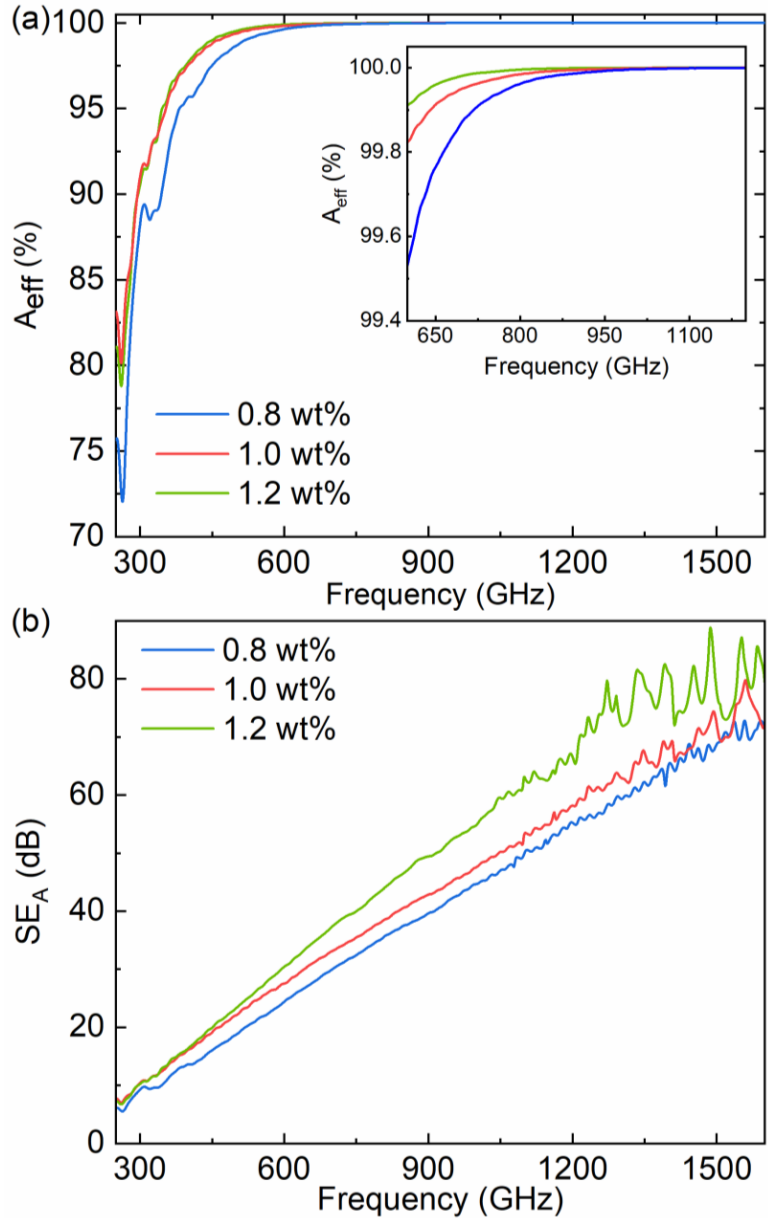

**Figure 5:** (a) Effective absorption coefficient, $A_{eff}$, of the epoxy composites with graphene as a function of frequency. The inset in (a) shows the same graph in the frequency range of $600\,\text{GHz} \leq f \leq 1200\,\text{GHz}$. Note that at frequencies $f \geq 950\,\text{GHz}$, the effective absorption reaches ~100%, indicating that the remaining fraction of the incident EM waves after reflection at the interface are completely absorbed by the composites. (b) Frequency-dependent shielding effectiveness of the composites by the absorption, $SE_A$.

### ▪ CONCLUSIONS

In conclusions, we have demonstrated that polymer composites with the low loading of graphene, below $1.2$ wt.%, are efficient as electromagnetic absorbers in the THz frequency range. The epoxy-based graphene composites were tested up to 4 THz using the Terahertz-Time Domain Spectroscopy system. It was found that the graphene composites with 1 mm thickness have the total shielding effectiveness of $\sim85$ dB at the frequency $f = 1.6$ THz. The THz radiation is mostly blocked by absorption rather than reflection. This is different from many other materials and composites used for EMI shielding. We have also observed changes in the transmission coefficient and the shielding effectiveness, which can be attributed to the frequency-dependent percolation effect. The efficiency of the THz radiation shielding by the lightweight, electrically insulating composites, increases with increasing frequency. We argue that even the thin-film or spray coatings of graphene composites with the thickness in the few-hundred-micrometer range can be sufficient for blocking THz radiation in many practical applications.

### ▪ METHODS

**Composite Preparation:** The composite samples were prepared by mixing graphene flakes with epoxy using a high-shear speed mixer (Flacktek Inc., U.S.A. with the in-house designed elements[71]) at 800 rpm and 2000 rpm each for 5 minutes. The mixture was vacuumed for 30 minutes. After that time, the curing agent (Allied High-Tech Products, Inc., U.S.A.) was added in the mass ratio of $12:100$ with respect to the epoxy resin. The compound was mixed and vacuumed one more time and left in the oven for $\sim2$ hours at 70 ℃ in order to cure and solidify.

### ■ ASSOCIED CONTENT

### ■ AUTHOR INFORMATION

**Corresponding Author**

\* Email: balandin@ece.ucr.edu (A.B.B.).

ORCID

Zahra Barani Beiranvand: 0000-0002-4850-0675

Fariborz Kargar: 0000-0003-2192-2023

Alexander A. Balandin: 0000-0002-9944-7894

## ACKNOWLEDGEMENTS

The work at UC Riverside was supported, in part, by the by the Office of Technology Partnerships (OTP), University of California *via* the Proof of Concept (POC) project "Graphene Thermal Interface Materials." This work was also supported by CENTERA Laboratories in frame of the International Research Agendas program for the Foundation for Polish Sciences co-financed by the European Union under the European Regional Development Fund (No. MAB/2018/9).

## CONTRIBUTIONS

A.A.B. and S.R. conceived the idea of the study, coordinated the project, and contributed to the EM data analysis. Z.B. prepared the composites, performed materials characterization and assisted with the EM data analysis. K.S. and Y.Y. performed the EM shielding measurements and contributed to data analysis. F.K. contributed to the sample preparation and EM data analysis. All authors contributed to writing and editing of the manuscript.